\documentstyle[aps,twocolumn]{revtex} 
\begin{document}  
\input{psfig}
\draft
\twocolumn[\hsize\textwidth\columnwidth\hsize\csname
@twocolumnfalse\endcsname
\title{Shape of the $^8$B alpha and neutrino spectra}
\author{C.E. Ortiz$^1$, A. Garc\'{\i}a$^{1,2}$, R. A. Waltz$^1$, 
M. Bhattacharya$^{1,a}$ and A. K. Komives$^1$}
\address{$^1$ University of Notre Dame, Notre Dame, Indiana 46556}
\address{$^2$ Lawrence Berkeley National Lab, Berkeley, California 94720}
\maketitle
\begin{abstract}                
The $\beta$-delayed $\alpha$ spectrum from the decay 
of $^8$B has been measured with a setup that minimized
systematic uncertainties that affected previous measurements.
Consequently the deduced neutrino spectrum
presents much smaller uncertainties than the previous
recommendation \cite{ba:96}.
The $^8$B $\nu$ spectrum is found to be harder than previously 
recommended with  about (10-20)\%
more neutrinos at energies between 12-14 MeV.
The efficiencies of the $^{37}$Cl, $^{71}$Ga, $^{40}$Ar,
and SuperKamioka detectors are respectively, 3.6\%, 1.4\%, 5.7\% and
1.8\% larger than previously thought.
\end{abstract}
\vspace{0.5cm}]
%
%
\renewcommand{\thefootnote}{\alph{footnote}}
\footnotetext[1]{Present address: Nuclear Physics Laboratory, 
University of Washington, Seattle, Washington 98195}
\narrowtext
The solar neutrino detectors SuperKamiokande, SNO, and ICARUS, are 
sensitive primarily to neutrinos from the decay:
$^8{\rm B} \rightarrow ~^8{\rm Be} + e^+ ~+\nu_e$.
The expected differences between the shape of this neutrino spectrum
in the laboratory and in the sun are small 
\cite{ba:91}. Hence any observed difference between the shape of this spectrum 
as measured in the laboratory compared to that measured by solar neutrino 
detectors would imply non-standard physics. For example, when the spectrum from 
SuperKamiokande is compared to the expected spectrum based on laboratory 
measurements\cite{ba:96}, one observes not 
just an overall reduction in the number of neutrinos but also a distortion 
of the spectrum,{\em i.e.}, the reduction is not as severe for the high-energy 
end of the spectrum (between 12 to 14 MeV) as it is for lower energies.
This has motivated several authors to try to find possible explanations
ranging from the physical
(hep neutrinos \cite{ro:99,BK:98,sch:99}, electron capture \cite{vil:99})
to the systematic (energy calibration uncertainties 
in SuperKamiokande \cite{bks:99}).

The $\beta^+$ decay of $^8$B$(J^\pi = 2^+)$ is dominated by a transition 
to a state at $E_x \approx 3$ MeV $(J^\pi = 2^+)$ with a width of 
$\approx 1.5$ MeV. Because this is a broad state, its interference with other 
$2^+$ states at higher energies affects the spectrum, and the final state 
distribution has to be determined experimentally. There have been 4 previous 
measurements of the $\alpha$ spectrum:
one by Farmer and Class (FC)\cite{fa:60},
one by DeBraeckeleer and Wright (DBW) (not published),
and two by Wilkinson and Alburger,
one with a thick catcher foil (WA1) 
and one with a thin catcher foil (WA2)\cite{WA:71}.
Because statistics are highest at the peak of the 
spectrum, Bahcall {\em et al.} \cite{ba:96} compared the 
different spectra by varying the energy calibration offset.
They showed that in order to get the best
agreement with the $\beta^+$ 
spectrum of Napolitano and Freedman\cite{na:87} the $\alpha$ 
spectra had to be shifted as follows: 
FC  by $\approx -85$ keV; 
DBW by $\approx +75$ keV; 
WA2 by $\approx +65$ keV; and 
WA1 by $\approx +25$ keV.                                                       
The high statistics measurement of Wilkinson and Alburger was 
motivated to search for Second Class Currents\cite{WA:71} by comparing 
the delayed-$\alpha$ spectra from $^8$Li and $^8$B decay. Consequently 
their measurement was optimized towards studying the {\em relative} 
differences between the two spectra and not their {\em absolute} shape.
Given the importance of an accurate knowledge of the shape
of the $^8$B neutrino spectrum, this state of affairs is
unsatisfactory.

\begin{figure}[p]
\psfig{figure=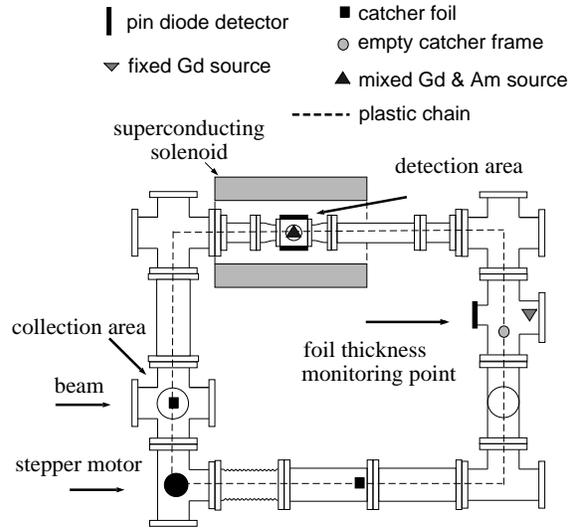,width=7.5cm,height=7.0cm,angle=0}
\caption{Overhead view of the experimental setup.}
\label{fig:setup}
\end{figure}

A common problem encountered by the previous measurements of the
delayed-$\alpha$ spectrum is the energy summing of the $\alpha$'s
with the preceding $\beta^+$'s, resulting in a distortion of the spectrum. 
In order to minimize this effect previous authors used 
small-solid-angle detectors. In addition, measuring the spectrum in 
singles, as was done by previous authors, entails subtracting 
low-energy $\beta^+$ backgrounds and possible events originating 
from $^8$B's implanted in the frame of the catcher foil,
correcting for shifts in the $\alpha$ energy due to the recoiling nucleus 
and $\alpha$-energy losses at different depths in the catcher foil,
all of which can introduce systematic distortions if not properly 
accounted for. Finally, all previous experiments had to be 
interrupted to perform detector energy calibrations.

In this paper the results from
an $\alpha$ spectrum measurement,
that overcame all of the difficulties discussed above,
are presented.

Figure~\ref{fig:setup} shows a top view of the setup. Catcher foils 
(20 $\mu$g/cm$^2$ of $^{12}$C) mounted on Al frames attached to a 
chain were transported through three stations by a computer-controlled 
stepper motor. At the first station a 5 MeV $^3$He beam (typically of  
$\sim 0.25$  p$\mu$A) impinged upon a target of $600$ to $700$ 
$\mu {\rm g/cm}^2$ of 95\% enriched $^6$LiF evaporated onto a $10$ 
$\mu {\rm g/cm}^2$ Carbon foil. The beam energy was chosen to optimize the 
production at the target. Because the $^6$LiF targets lasted only for 
a few hours of bombardment, several targets were mounted in a 
vertical-motion ladder, which allowed for quick replacement 
without having to break the vacuum. The radioactivity was 
collimated to make a 0.79 cm diameter circular spot on the catcher foils, 
while the hole in the Al frames for the catcher foils was 1.27 cm in diameter.
The second station had two Hammamatsu Si PIN detectors 
with 256 mm$^2$ of effective area (after masking their edges) located 
$3.24 \pm 0.16$ cm from either side of the catcher foil. 
The counting chamber 
was placed inside the bore of a super-conducting solenoid which produced
a 3.5-Tesla magnetic field perpendicular to the line joining the centers of 
the detectors. The magnetic field channeled the $\beta$'s away from the 
detectors. The third station was dedicated to monitor the catcher foil 
thickness by measuring the energy loss of $^{148}$Gd $\alpha$ particles
passing through the catcher foils. The catcher foil thicknesses were 
monitored throughout the data taking. During the course of the run the 
catcher foils grew thicker due to a combination of sputtering from the target 
and carbon build up on the beam spot due to imperfect vacuum, and  
were replaced when their thickness exceeded 30 $\mu$g/cm$^2$.

The data was taken by repeating cycles which consisted of four stages. 
In the first stage while radioactivity was being collected on the first 
catcher foil, both of the detectors at the counting station faced mixed 
$^{148}$Gd ($E_\alpha \approx 3183$ keV) and 
$^{241}$Am ($E_\alpha \approx ~5443,~5486~{\rm and}~5499$, keV)
thin $\alpha$ sources. At the end of the collection the chain and sprocket 
system (see Fig.~\ref{fig:setup}) was rotated to the next position.
The distance of $\approx 0.9$ m from the loading to the
counting station was covered in $\approx 0.9$ s, so the
radioactivity having a half life of less than a second could
be counted efficiently. In the second stage the first catcher 
foil would be counted by the detectors at the counting station 
while the second one was being loaded. In the third stage the 
thickness of the first catcher foil was monitored at the 
thickness-monitoring station, while the second catcher foil was 
being counted at the counting station and an empty catcher-foil frame 
was at the collection area. The purpose of the empty frame was to 
monitor the radioactivity implanted in the aluminum frame.
In the fourth and final stage the thickness of the second catcher 
foil was monitored while the empty catcher foil frame was at the 
detection area. The chain would then be rotated in the reverse 
direction until the system was back to the original position. 
Each collection-counting stage took $\approx 2.5$ seconds.
The incident beam was interrupted by a Faraday cup, located 
$\sim$ 5 meters upstream of the system, every time the chain 
moved, to avoid depositing radioactivity in places other than the 
center of the catcher foils.

The primary advantages of the experimental setup in light
of the discussion above are:
\begin{enumerate}

\item As the detectors were placed in a $3.5$-Tesla magnetic field,
the positrons from the decay of $^8$B  could not reach the
detectors while the delayed $\alpha$'s suffered little deflection.
This allowed the detection of $\alpha-\alpha$ coincidences
without $\beta-\alpha$ contamination. This is a powerful
tool to reject backgrounds and avoid counting $\alpha$'s
coming from radioactivity implanted in the Al frame.
The coincidence summed spectra are also free of 
any recoil-broadening.

\item Energy calibrations of the detectors were performed during each 
cycle, without having to break the vacuum or un-bias the detectors
to introduce calibration sources. 

\item The foils thicknesses were monitored throughout the course
of the experiment.

\end{enumerate}

The signals were shaped using ORTEC 142A pre-amplifiers and ORTEC 572 
amplifiers and digitized using an ORTEC 413A ADC. The trigger 
was defined as a hit in any one of the three detectors.
 
The recorded pulse heights were corrected for energy losses in 
the catcher foil and detector dead layers on an event-by-event basis.
The largest corrections (at $E_\alpha = 0.5$ MeV) were $\approx 25$ keV
and $\approx 15$ keV for energy losses in the catcher foil and detector 
dead layer respectively. The detectors dead layer was measured before 
and after the data taking by mounting the detectors on a rotating jig which 
allowed a measurement of the energy deposited by a $^{148}$Gd $\alpha$
in the detector as a function of the angle between the normal
to the source and the normal to the detector. The dead layers were 
assumed to be 100\% Si and were found to be $9 \pm 2$ $\mu {\rm g/cm}^2$ 
for both detectors. 
The $\alpha$'s summed energy was also corrected for the effect of
the recoiling $^8$Be nucleus. The largest
correction for this effect (at $E_{\alpha 1} + E_{\alpha 2} \approx 1.0$ MeV)
was $\approx 8$ keV.
A Monte Carlo simulation was used to determine 
the average amount of energy each $\alpha$ would lose in the catcher 
foil and in the detector dead layers
as well as the average velocity of the recoiling $^8$Be nucleus
as a function of the $\alpha$'s energy.

\begin{figure}[p]
\psfig{figure=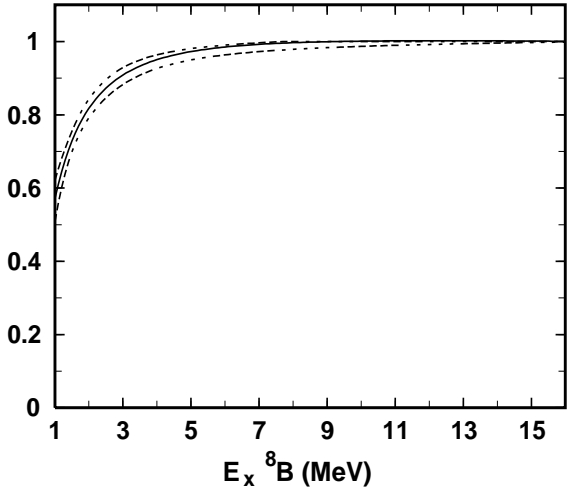,width=7.5cm,height=6.5cm,angle=0}
\caption{Average efficiency (solid)
and systematic uncertainties (dashed) for detecting a pair of
$\alpha$ particles vs. excitation energy in $^8$Be.
The upper dashed line represents the uncertainty in the efficiency
due to the equilibrium thickness and the source's diameter.  The lower dashed
line includes these uncertainties as well as those 
due to the source's position (misalignment can only reduce the efficiency).}
\label{fig:field_eff}
\end{figure}

One disadvantage of the experimental setup is the fact that the 
energy dependence of the $\alpha$ efficiency
is magnified by the presence of the magnetic field.
Figure~\ref{fig:field_eff} shows the average efficiency for 
detecting a pair of $\alpha$ particles as a function of 
the $^8$Be excitation energy. 
The efficiency was calculated using a Monte Carlo code that took into 
account the effect of the magnetic field, the charge distributions of 
the $\alpha$ particles coming out of the Carbon foil, and the  
source size and position. The source size was determined by the 
collimators and confirmed by visual inspection of the catcher foils.
The source position was monitored and
found to be extremely stable during data taking. The ratio of coincidences 
to singles events turned out to be a powerful tool in constraining 
systematic uncertainties. Using both sources of information
the  source position was constrained to $\pm 0.16$ cm in all three spatial
directions, and the diameter of the source spot to 
$0.79 \pm 0.16$ cm. 
The {\em equilibrium} charge state distribution
of $\alpha$ particles for a given energy was interpolated 
from the measured values quoted in Ref. \cite{sk:allison}. The minimum 
thickness needed to reach equilibrium, $x_{\rm min}$, was estimated to be 
$\approx 5 \mu$g/cm$^2$. To account for possible uncertainties in this estimate
of the charge state distribution, calculations were performed assuming 
$x_{\rm min}= 1$, 5, and 10 $\mu$g/cm$^2$; but this contribution to
the uncertainty in the efficiency was found to be negligible.
For all cases the charge distribution was assumed to
follow a straight line 
between the equilibrium point, $x_{\rm min}$, and the edge of the foil, 
from where an $\alpha$ is emitted in the $q=+2$ state.

The data was fit under 8 different conditions.
Under condition A each detector's dead
layer was assumed to be 9 $\mu$g/cm$^2$ and each catcher foil
was corrected for changes in thickness over time.
Under condition B each detector's dead layer was assumed to
be 7 $\mu$g/cm$^2$ and each catcher foil was assumed to
be a constant 20 $\mu$g/cm$^2$.
Under condition C each detector's dead layer was assumed to
be 11 $\mu$g/cm$^2$ and each catcher foil was assumed to
be a constant 30 $\mu$g/cm$^2$.
Condition B underestimates the amount of energy
lost in the dead layers and catcher foil, while condition
C overestimates this quantity.
Condition D was the same as A except that no line-shape correction
was made. 
Three additional conditions, E$_x$, E$_y$, and E$_z$,
were the same as A except that efficiency used was
for a source displaced
by 0.16 cm in the $x$, $y$, and $z$ spatial directions
respectively.
Condition E$_d$ assumed an efficiency for a source
diameter 0.16 cm larger than expected.
The statistical errors to the $\alpha$ spectrum
were calculated as the square root
of the number of raw counts divided by the relative efficiency.
The systematic errors to the $\alpha$ spectrum were calculated
by adding in quadrature the differences between the R-matrix
fit under condition A to those under the 7 other conditions
and condition A offset by 8.5 keV (the error in the calibration). 
A table with the results can be down-loaded~\cite{www:tables}.
\begin{figure}[p]
\psfig{figure=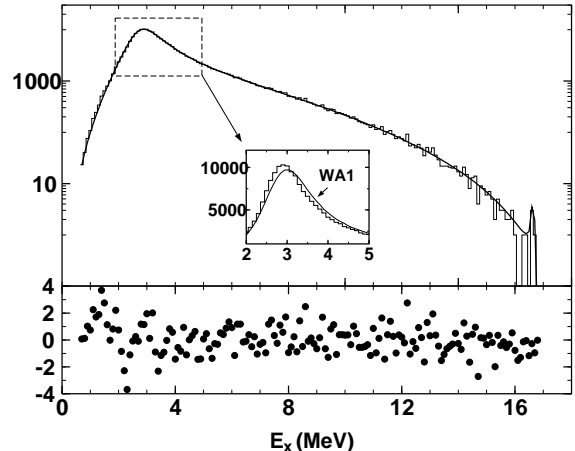,width=7.5cm,height=6.0cm,angle=0}
\caption{Top panel: $\alpha$ spectrum and R-matrix fit.
The insert compares this spectrum (stair step) to the WA1 spectrum
(solid) normalized to have the same
number of counts as this $\alpha$ spectrum.
Bottom panel: residuals.}
\label{fig:alp_spec}
\end{figure}
Figure~\ref{fig:alp_spec} shows an R-matrix fit to 
our total $\alpha$ spectrum,
which was performed following the prescriptions of Refs.~\cite{WB:86,BK:89}
and limiting the fits to contain 4 resonances and not including an 8-MeV 
intruder state. The data is well described with 
$\chi^2/\nu \approx 1.1$ with $\nu \sim 160$, 
while the fits to the WA1 and WA2 spectra 
yield $\chi^2/\nu \approx $1.9 and 2.2, respectively, 
assuming only statistical errors.
The inset shows a comparison between these data (histogram) and
the WA1 spectrum (continuous line). The present spectrum peaks at a lower
excitation energy.

The intensity of the second forbidden
transition to the $0^+$ ground state 
was estimated using the known
$\Gamma_\gamma$ widths\cite{de:95} of the decays from the two states at
$E_x \approx 16$ MeV and the Conserved
Vector Current hypothesis
to extract the matrix element. A rough estimate
is that this should be $\approx 10$ orders of magnitude
less intense than the allowed decay. 

 From the R-matrix fit to the total $\alpha$ spectrum
a $\beta^+$ spectrum was deduced and
compared to the data of Ref. \cite{na:87}.  Allowing
the momentum calibration of Ref. \cite{na:87} to be
offset by a constant value, $b$, a minimum
$\chi^2$ of $31.8$ was found for 33 data points
and $b\approx 70\pm20$ keV/c.  Thus this $\alpha$ spectrum
predicts a harder $\beta^+$ spectrum than
the calibrated spectrum of Ref. \cite{na:87}. In this sense,
this $\alpha$ spectrum is in 
rough agreement with the WA2 and DBW spectra.

Using the {\em distribution of strength} from the R-matrix fits to
foils 1 and 2 the $^8$B neutrino spectrum was deduced. Small 
radiative and forbidden corrections were included
using the same formalism as 
in Ref.~\cite{ba:96}, except that more complete expressions 
were used for the forbidden corrections but the differences
were found to be negligible.
The best estimate of the $^{8}$B $\nu$ spectrum
was taken as the average of the spectra generated using the 
R-matrix parameters for foils 1 and 2 under condition A. 
\begin{figure}[p]
\psfig{figure=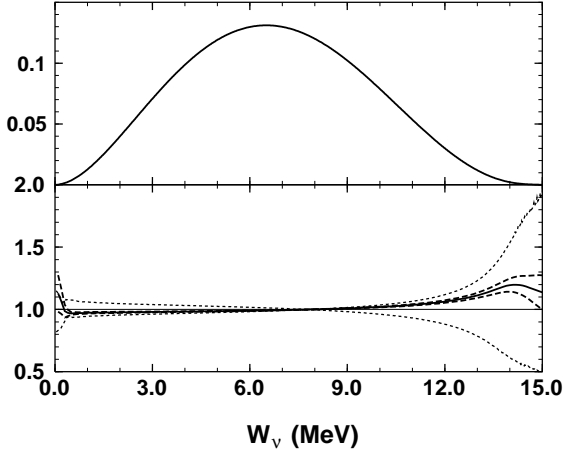,width=7.5cm,height=6.0cm,angle=0}
\caption{Top panel: neutrino spectrum.
Bottom panel: ratio between this neutrino spectrum and
that of Ref.~\protect\cite{ba:96}.
The dashed lines represent the uncertainties
in this spectrum while the dotted lines represent
the $3\sigma$ errors in Ref. \protect\cite{ba:96}.}
\label{fig:diff_pcal}
\end{figure}
\begin{figure}[p]
\psfig{figure=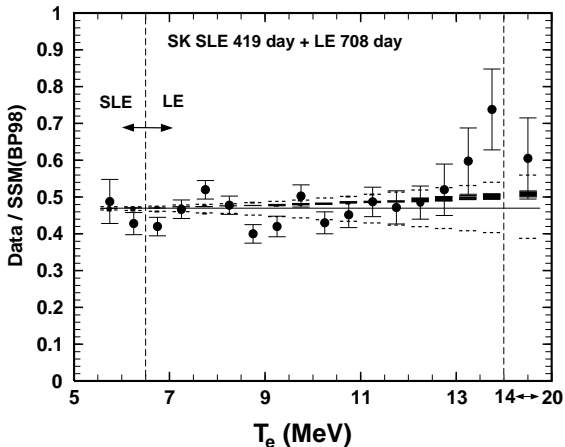,width=7.5cm,height=6.0cm,angle=0}
\caption{The 708-day spectrum of electrons at SuperK
divided by what is expected based on
the $\nu$ spectrum shape of Ref.~\protect\cite{ba:96},
and the solar model of Ref.~\protect\cite{bp:98} (BP98).
The data points and total errors were taken graphically
from Ref. \protect\cite{smy:99}.
Also plotted is 0.47 times \protect\cite{smy:99} the ratio
of the electron spectrum generated using this $\nu$
spectrum to that of Ref.~\protect\cite{ba:96}
with the SuperK
response function.  The width of the solid line represents
the uncertainty associated with this neutrino spectrum while
the dotted lines represent the $3\sigma$ errors to the
$\nu$ spectrum of Ref.~\protect\cite{ba:96}.}
\label{fig:correction}
\end{figure}

The ratio of this $\nu$ spectrum to the best recommendation of 
Ref. \cite{ba:96} is shown in Fig. \ref{fig:diff_pcal}.
Approximately $10-20$\% more neutrinos are found in the high-energy
end of the spectrum.
The uncertainties in the  $\nu$ spectrum were calculated in a 
similar way as explained above for the alpha spectrum except
one more condition was added.
The neutrino spectrum was
generated using the data
itself instead of an R-matrix fit.
Figure~\ref{fig:correction} shows 
the implications of these findings with respect 
to the SuperKamiokande data.
This spectrum implies a correction to Ref.~\cite{ba:96}'s 
recommendation that ranges from 
$\approx 0$\% at $T_e = 5$ MeV to $\approx 8$\% at the endpoint. 

The efficiencies of the $^{37}$Cl, $^{71}$Ga, $^{40}$Ar,
and SuperKamioka detectors are respectively, 3.6\%, 1.4\%, 5.7\% and
1.8\% larger than previously thought.

\acknowledgments
We thank J.J. Kolata, F. Bechetti, D. Peterson, and P. Santi,
for help during the early stages of this experiment,
J. Napolitano for sending us the $\beta^+$ spectrum,
and J.F. Beacom, R.G.H. Robertson and S.J. Freedman 
for illuminating comments.
AG thanks the National Institute for Nuclear Theory at Seattle
for hosting during the summer of 1999.
%
%

%
%
%
%
%

\begin{references}
%
\bibitem{ba:96} J.N. Bahcall, E. Lisi, D.E. Alburger,
L. De Braeckeleer, S.J. Freedman and J. Napolitano, 
Phys. Rev. C {\bf 54}, 411 (1996).
%
\bibitem{ba:91} J.N. Bahcall, 
Phys. Rev. D {\bf 44}, 1644 (1991).
%
\bibitem{ro:99}R.G.H. Robertson, 
Presented at Lepton-Photon 99 Conference, Stanford,
Aug. 9-14, 1999; to be published in the Proceedings;
LANL hep-ex/0001034v2.
%
\bibitem{BK:98}J.N. Bahcall and P.I. Krastev,
Phys. Lett. B {\bf 436}, 243 (1998).
%
\bibitem{sch:99} R. Schiavilla,
Bull. Am. Phys. Soc. {\bf 44}, 28 (1999).
%
\bibitem{vil:99} F.L. Villante,
Phys. Lett. B {\bf 460}, 437 (1999).
%
\bibitem{bks:99}J.N. Bahcall, P.I. Krastev, and A. Yu. Smirnov,
Phys. Rev. D {\bf 60}, 093001 (1999)
%
\bibitem{fa:60} B.J. Farmer and C.M. Class,
Nucl. Phys. {\bf 15}, 626 (1960).
%
\bibitem{WA:71} D.H. Wilkinson and D.E. Alburger, 
Phys. Rev. Lett. {\bf 26}, 1127 (1971).
%
\bibitem{na:87}J. Napolitano and S.J. Freedman,
Phys. Rev. C {\bf 36}, 298 (1987).
%
\bibitem{sk:allison}S.K. Allison, Rev. Mod. Phys.
{\bf 30}, 1137 (1958).
%
\bibitem{www:tables}{Tables containing our alpha and deduced neutrino
spectra can be found at {\bf www.nd.edu/$\sim$nsl/BeyondSM/boron8/tables}}.
%
\bibitem{WB:86} E.K. Warburton, 
Phys. Rev. C {\bf 33}, 303 (1986) and references therein.
%
\bibitem{BK:89} F.C. Barker,
Aust. J. Phys. {\bf 42} , 25 (1989)
%
\bibitem{de:95} DeBraeckeleer {\em et al.},
Phys. Rev. C {\bf 51}, 2778 (1995).
%
\bibitem{bp:98}J.N. Bahcall, S. Basu, and M. Pinsonneault,
Physics Letters B {\bf 433}, 1 (1998).
%
\bibitem{smy:99} M.B. Smy, hep-ex/9903034 
%
\end{references}
\end{document}